\documentclass[prb,twocolumn,showpacs]{revtex4-1}
\usepackage{graphicx}
\usepackage{times}
\usepackage{color}
\input pdfcolor.tex
\usepackage{graphicx}
\usepackage{epsfig,color}
\usepackage{amsfonts,amssymb}
\usepackage{amsmath}
\usepackage{color}
\usepackage{epstopdf}
\definecolor{refcolor}{rgb}{1.0,0.0,0.0}

\newcommand{\be}{\begin{equation}}
\newcommand{\ee}{\end{equation}}   
\newcommand{\bea}{\begin{eqnarray}}
\newcommand{\eea}{\end{eqnarray}}
\newcommand{\ba}{\begin{array}}
\newcommand{\ea}{\end{array}}

\newcommand{\phrl}[1]{Phys.~Rev.~Lett., {\bf #1}}
\newcommand{\phrb}[1]{Phys.~Rev.~B, {\bf #1}}

\newcommand{\jpsj}[1]{J.~Phys.~Soc.~Jpn.{\bf #1}}

\begin{document}

\title{Drude weight anisotropy in the doped iron pnictides:
the primary role of\\
 orbital weight redistribution along 
the reconstructed Fermi surfaces}

\date{\today}
\author{Dheeraj Kumar Singh$^{1,2,3}$ and Pinaki Majumdar$^{3}$}

\affiliation{$^1$Department of Physics, POSTECH, Pohang, Gyeongbuk 790-784, Korea}
 
\affiliation{$^2$Asia Pacific Center for Theoretical Physics, Pohang, Gyeongbuk 790-784, Korea}
 
\affiliation{$^3$Harish-Chandra Research Institute, HBNI, 
Chhatnag Road, Jhunsi, Allahabad 211019, India}

\begin{abstract}
We explore the anisotropy in low frequency conductivity as a function of 
hole doping in multiorbital models of pnictides by analyzing the Drude 
weight in the $x$  and $y$ directions of a ($\pi$, 0) spin-density wave 
state. A reduction in the conductivity anisotropy with increased hole 
doping, and subsequent sign reversal, both observed in experiments, is 
found to be a common trend in these models. This behavior arises due to 
the interplay of low energy orbitally-resolved density of states and the 
geometrical features of the Fermi surface. An understanding of anisotropy 
in the electron doped regime, however, would require additional ingredients.
\end{abstract}


\maketitle

\section{Introduction}
The anisotropic electronic properties of the iron pnictides,
in both the  collinear spin-density wave (SDW) state and 
the nematic state, have attracted considerable attention. 
Evidence of anisotropy is visible 
in angle-resolved photoemission spectroscopy~\cite{yi,shimojima1} (ARPES), nuclear magnetic resonance 
\cite{fu} (NMR), scanning tunneling microscopy~\cite{chuang}, 
and transport measurements~\cite{chu,tanatar,blomberg}. According to the 
ARPES measurements, a significant energy splitting between $d_{xz}$ and 
$d_{yz}$ orbitals is observed below the tetragonal-to-orthorhombic
transition, which may either precede~\cite{nandi} 
the SDW transition
or occur simultaneously~\cite{rotter}, and states at the Fermi level
have dominantly $d_{xz}$ character~\cite{shimojima2}.
In-plane anisotropy is observed also in the optical 
spectra~\cite{nakajima} up to a photonic energy of 2eV.

The easiest to measure is transport anisotropy. 
Detwinned crystals show
larger conductivity, surprisingly, in the antiferromagnetic
direction compared to the ferromagnetic 
direction~\cite{chu, tanatar, ying}. The ratio of resistivities 
can be as large as $1.5$ in
the undoped parent compound. Similar conductivity 
anisotropy has also been observed in the nematic 
state~\cite{ying,blomberg}.

Theories have tried to address the anisotropy. 
In the collinear SDW state of the
undoped compound  
multiorbital models have been studied 
employing (i)~a 
combination~\cite{yin}
of the local density approximation (LDA) 
and dynamical mean field theory (DMFT), as well as 
(ii)~mean-field methods~\cite{sugimoto1,zhang, valenzuela}.
According to these studies, a significant difference in the
density of states (DOS) of $d_{xz}$ and $d_{yz}$
orbitals at the Fermi level may be responsible for  
the counter-intuitive anisotropy in the conductivity. Suppression 
of the DOS of $d_{yz}$ orbital \cite{daghofer1} 
at the Fermi level, compared to that of $d_{xz}$,  
in the ($\pi, 0$)-SDW state results in a larger 
conductivity along the $x$-direction since a major
contribution to the Drude weight comes from the interorbital 
$d_{xz}$ to $d_{xy}$ hopping~\cite{zhang}. 
Conductivity in the $y$-direction, 
facilitated primarily by hopping between
$d_{yz}$ and $d_{xy}$, is smaller. Another 
study~\cite{valenzuela} emphasizes the role of Fermi 
surface `reconstruction'. 
What about doped systems?

Experiments reveal that the conductivity anisotropy
decreases and changes sign as hole doping increases~\cite{blomberg} while it grows on electron doping~\cite{chu, ying}. 
In order to describe the doping dependence the
role of doping induced disorder has been invoked.  
It was shown that disorder with isotropic impurity potential 
can lead to elongated magnetic droplets in the nematic 
state, with anisotropy increasing as one approaches 
the SDW transition~\cite{gastiasoro,timm,wang,dheeraj}. Deep inside the SDW state, 
the disorder has been suggested to be 
important particularly close to zero doping,
where the Dirac 
point is situated near the Fermi level ~\cite{sugimoto}.

The role of both magnetic and non-magnetic 
disorder has indeed been explored experimentally.
In the tetragonal state
subjected to external strain, electron-doped 
samples having different
residual resistivity 
show similar resistivity anisotropies~\cite{kuo}. Studies 
on  electron- and hole-doped samples, and the presence of
magnetic impurities, drew a similar conclusion. 
This suggests that while disorder controls the
magnitude of the resistivity, the anisotropy is 
more a consequence of the bandstructure 
in the doped system~\cite{kobyashi}.

To conclusively settle this issue, we 
examined the Drude weight both 
in (a)~the
($\pi, 0$)-SDW state of the three-orbital model proposed by 
Daghofer \textit{et al.}~\cite{daghofer} and (b)~the more 
sophisticated five-orbital models of Graser \textit{et al.}~
\cite{graser} and Ikeda \textit{et al.}~\cite{ikeda}.
We find the following: 
(i)~Despite considerable differences 
all these models exhibit the robust common 
feature that conductivity anisotropy decreases with 
hole doping, and subsequently reverses it's sign
- all consistent with experiments. 
(ii)~We explain this in terms of the 
orbitally resolved weight along the Fermi surface, the 
hopping anisotropy, 
and shape of the Fermi surface. 
(iii)~In the electron-doped region, 
the ratio of Drude weights 
along ferromagnetic and antiferromagnetic direction approaches
$\sim 1$ asymptotically. Thus, 
experimental trend observed on electron doping
is not captured within this approach, suggesting additional
effects at play.
\section{Theory}
We consider a multiorbital model Hamiltonian with the kinetic
part defined as  
\begin{equation}
\mathcal{H}_0 = -\sum_{{\bf i},{\bf j}}^{\mu,\nu, \sigma}
t_{{\bf i}{\bf j}}^{\mu\nu} d_{{\bf i}\mu\sigma}^\dagger d_{{\bf j}\nu\sigma}
\end{equation} 
where the operator $d_{{\bf i} \mu \sigma}^\dagger$ ($d_{{\bf i}
\mu \sigma}$) creates (destroys) an electron in the $\mu$-th orbital
of site ${\bf i}$ with spin $\sigma$, and $t_{{\bf i}{\bf j}}^{\mu\nu}$
are the hopping elements from orbital $\mu$ and $\nu$ of sites ${\bf i}$
and ${\bf j}$, respectively. The orbitals $\mu$ and $\nu$ belong to the set 
of five $d$-orbitals $d_{xz}$, $d_{yz}$, $d_{xy}$, $d_{x^2-y^2}$, and 
$d_{3z^2-r^2}$ depending on the model.

The interaction part of the Hamiltonian is given by
\begin{eqnarray}
\mathcal{H}_{int} &=& U \sum_{{\bf i},\mu} n_{{\bf i}\mu
\uparrow} n_{{\bf i}\mu \downarrow} + (U' - \frac{J}{2})
\sum_{{\bf i}, \mu<\nu, \sigma \sigma^{\prime}} n_{{\bf i} 
\mu \sigma} n_{{\bf i} \nu \sigma^{\prime}} 
\cr
&-& 2 J \sum_{\bf i}^{\mu<\nu} {\bf{S_{{\bf i} \mu}}} 
\cdot {\bf{S_{{\bf i} \nu}}} 
+ J \sum_{\bf i,\sigma}^{\mu<\nu} 
d_{{\bf i} \mu \sigma}^{\dagger}d_{{\bf i} \mu 
\bar{\sigma}}^{\dagger}d_{{\bf i} \nu \bar{\sigma}}
d_{{\bf i} \nu \sigma} \nonumber
\label{int}
\end{eqnarray}
which includes the intraorbital (interorbital) Coulomb 
interaction term as the first (second) term. 
The third term describes the Hund’s coupling, and the 
fourth term represents the pair hopping energy. Rotation-invariant 
interaction is ensured provided that $U^{\prime}$ = $U$ - $2J$. 

The Hamiltonian after mean-field decoupling of the interaction 
term in the $(\pi, 0)$-SDW state is given in the two sublattice
basis by   
\begin{equation} 
\mathcal{H}_{mf} = \sum_{\bf k \sigma} \Psi^{\dagger}_{{\bf k} 
\sigma} (T_{{\bf k} \sigma} + M_{{\bf k} \sigma})
\Psi_{{\bf k} \sigma},
\end{equation}
where matrix elements $T_{{\bf k} \sigma}^{ll^{\prime}}$ and
$M^{ll^{\prime}}_{{\bf k} \sigma} = -s \sigma \Delta_{l^{\prime}} 
\delta^{ll^{\prime}} + \frac{5J - U}{2} n_{l^{\prime}} \delta^{ll^{\prime}}$
are due to the kinetic and interaction parts, respectively. 
$l$, $l^{\prime}$ $\in$ $s \otimes \mu $ with $s$ and
$\mu$ denoting a sublattice and an orbital, respectively. 
$s$ and $\sigma$ in front of $\Delta_{ll^{\prime}} \delta^{ll^{\prime}}$
take value 1 (-1) for A (B) sublattice and $\uparrow$-spin ($\downarrow$-spin),
respectively. The electron field operator and the exchange fields are 
defined as $\Psi^{\dagger}_{k} = (d^{\dagger}_{A{\bf k}1 
\uparrow},d^{\dagger}_{A{\bf k}2 \uparrow},. .. ,d^{\dagger}_
{B{\bf k}1 \uparrow},d^{\dagger}_{B{\bf k}2 \uparrow},...)$ and $2\Delta_{l} =
Um_{l} + J \sum_{l \ne l^{\prime}}m_{l^{\prime}}$, respectively. 
Charge density $n_l$ and magnetization 
$m_l$ in each of the orbitals are determined
self-consistently by diagonalizing the Hamiltonian. 

\begin{figure}[b]
\begin{center}
\psfig{figure=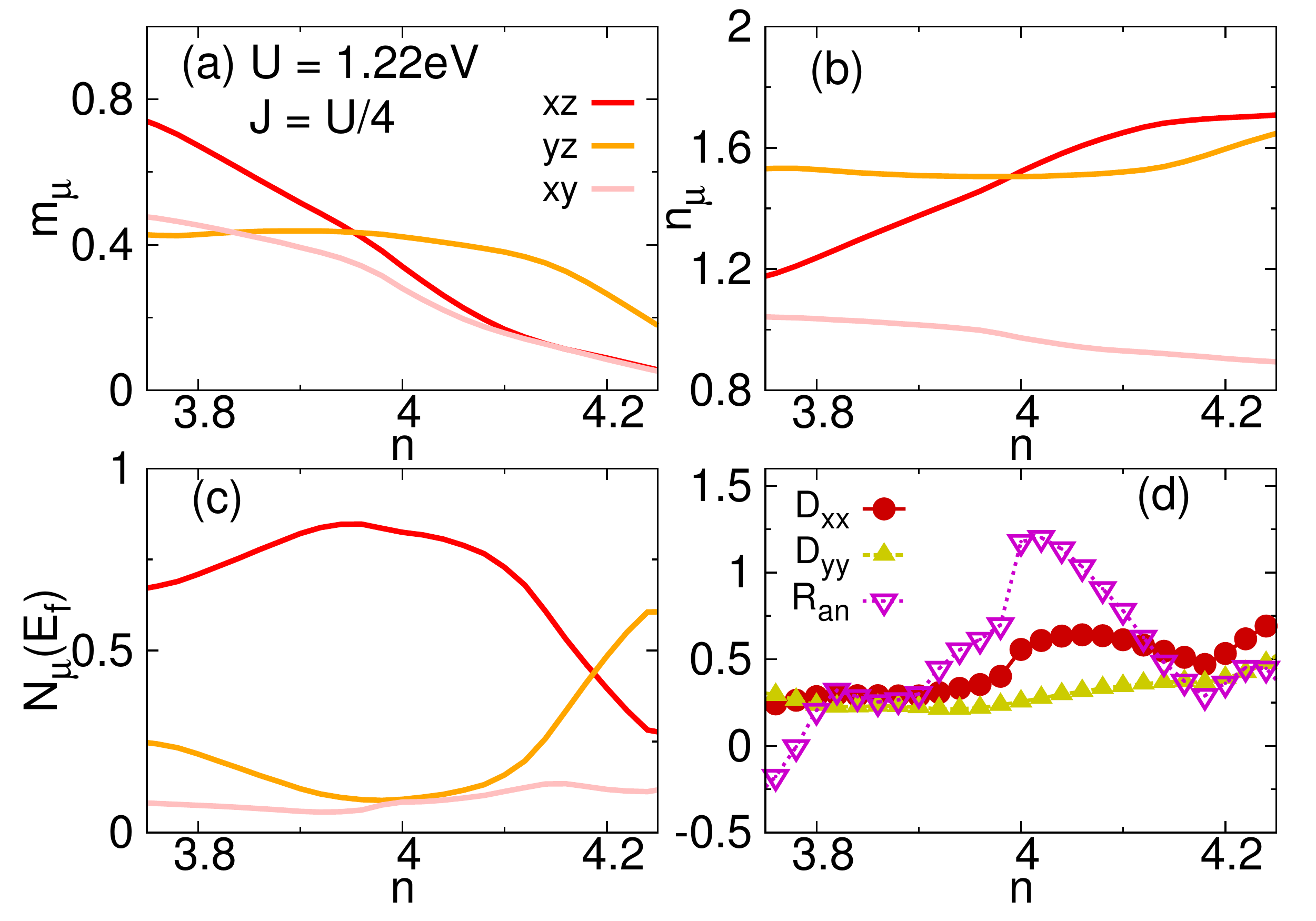,width=6.2cm,height=8.0cm,angle=-90}
\end{center}
\vspace{-4mm}
\caption{Doping dependence of the (a) magnetic moment,
(b) electronic density, and (c) DOS at the
Fermi level for each of the orbitals in the ($\pi, 0$)-SDW
state of the three-orbital model of Daghofer \textit{et al}..
(d) Drude weights $D_x$ and $D_y$ in the $x$-direction and $y$-direction, 
respectively. For convenience, the dimensionless 
quantity $R_{an} = D_x$/$D_y -1$, which provides a 
measure of the anisotropy, is also plotted.}
\label{daghofer1}
\end{figure}
\begin{figure}[t]
\begin{center}
\psfig{figure=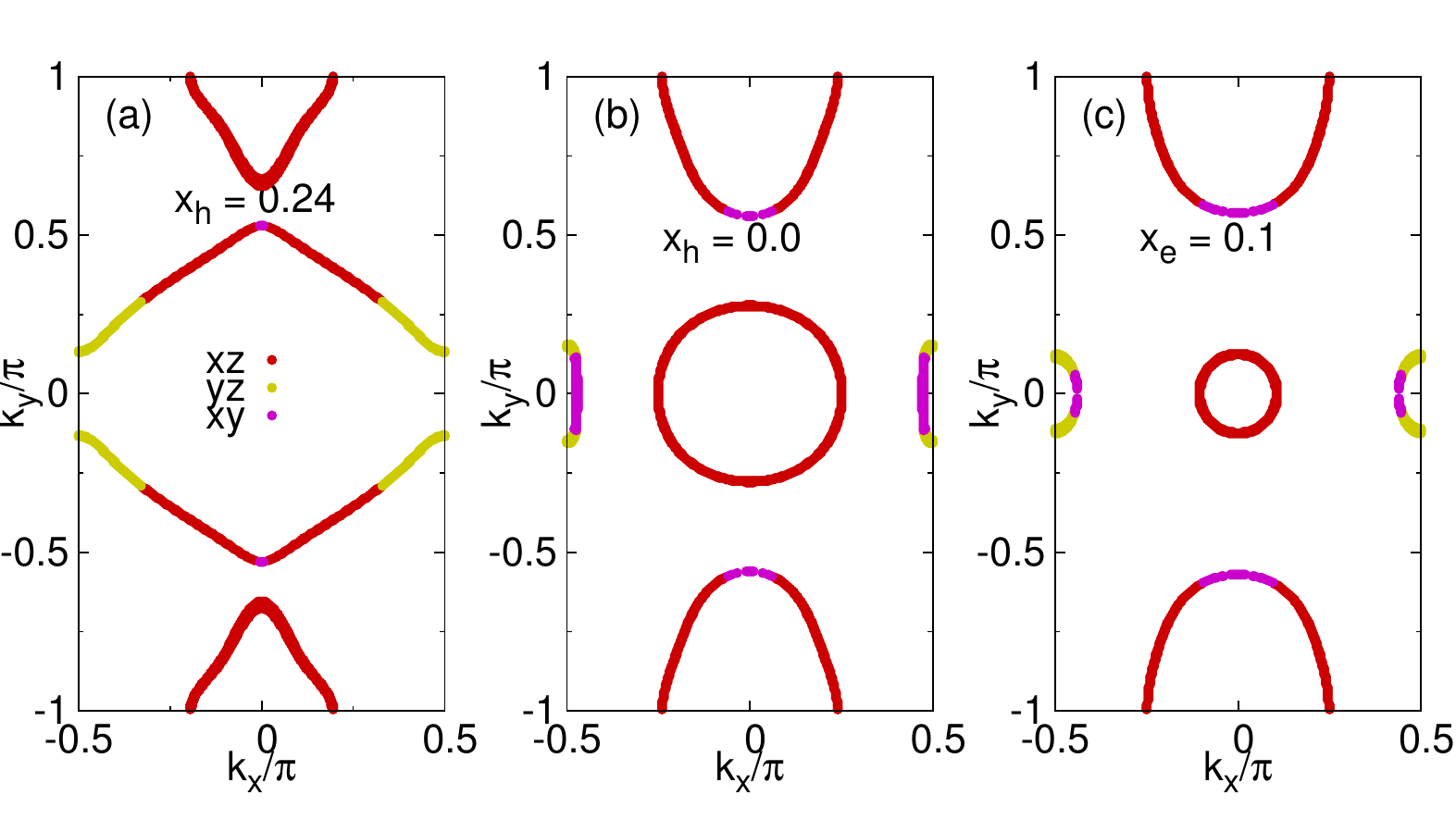,width=80mm,height=5.6cm,angle=0}
\end{center}
\vspace{-4mm}
\caption{Fermi surfaces in the ($\pi, 0$)-SDW state of
the model of Daghofer \textit{et. al.} for hole dopings
(a) $x_h = 0.24$, (b)  $x_h = 0.0$, and electron doping
(c)  $x_e = 0.1$ with the predominant orbital densities,
where $U = 1.22$eV.}
\label{daghofer2}
\end{figure}

The optical conductivity along $\alpha  = x,y$ given by 
\cite{dagotto,valenzuela1} 
\begin{eqnarray} \sigma_{\mathcal{\alpha}} &=& D_{\alpha} \delta{(\omega)}+
\frac {1}{N} \sum_{{\bf k}, n \ne n^{\prime}} 
\frac{ |j^{\alpha}_{nn^{\prime}}({\bf k})|^2}{E_{n^\prime {\bf k}} - 
E_{n {\bf k}}} \nonumber\\
&\times&\theta (-E_{n^\prime {\bf k}})\theta (E_{n {\bf k}}) 
\delta(\omega-E_{n^\prime {\bf k}}+E_{n {\bf k}})
\end{eqnarray}
with Drude weight  
\begin{eqnarray} \frac {D_{\alpha}}{2 \pi} &=& \frac{1}{2N}
\sum_{\bf k} T^{\alpha}_{nn} ({\bf k}) \theta (-E_{n {\bf k}}) -
\frac {1}{N} \sum_{{\bf k}, n \ne n^{\prime}} 
\frac{ |j^{\alpha}_{nn^{\prime}}({\bf k})|^2}{E_{n^\prime {\bf k}} - 
E_{n {\bf k}}} \nonumber\\
&\times&\theta (-E_{n^\prime {\bf k}})\theta (E_{n {\bf k}}).
\end{eqnarray}
$\theta$ is the step function, $E_{n{\bf k}}$ is the single 
particle energy in the ordered state with $n$ as a band index, and
\begin{eqnarray} T^{\alpha}_{nn} = \sum_{\mu \nu} T^{\alpha;\mu \nu}_{nn}
= \sum_{\mu \nu} \frac {\partial^2 
t_{\mu \nu} ({\bf k})}{\partial k^2_x}
a^*_{{\bf k} \mu n} a_{{\bf k}\nu n} 
\nonumber\\
j^{\mathcal{\alpha}}_{nn^{\prime}} = -\sum_{\mu \nu} j^{\alpha;\mu \nu}_{nn^{\prime}} 
= -\sum_{\mu \nu} \frac {\partial
t_{\mu \nu} ({\bf k})}{\partial k_x} 
a^*_{{\bf k} \mu n} a_{{\bf k}\nu n},
\end{eqnarray}
$a_{{\bf k} \mu  n}$ is the unitary matrix element 
between the orbital $\mu$ and band $n$ in ($\pi, 0$)-SDW state. To 
investigate the role of electronic reconstruction in the conductivity anisotropy, 
electron scattering by impurities or spin fluctuations can be assumed to be isotropic
for simplicity, which may not be the case in actuality~\cite{kemper}.
However, absence of anisotropy 
due to impurity scattering is also supported by experiments. Then,
$\delta$ function is approximated by Lorentzian with a small but non vanishing 
broadening parameter, which is same in both directions.

In this work we focus on the anisotropy in $\omega \rightarrow 0$ limit. To gain 
a better understanding of the origin of
anisotropy, it is useful to define the components of 
Drude weight as~\cite{zhang}
\begin{eqnarray}
 \frac {D^{\alpha\beta}_{l}}{2 \pi} &=& \frac{1}{2N}
\sum_{\bf k} T^{l;\alpha \beta}_{nn} ({\bf k}) \theta
(-E_{n {\bf k}}) \nonumber\\ 
&-& \frac {1}{N} \sum_{{\bf k}, n \ne n^{\prime}}  Re
\frac{ j^{x(y);\alpha {\beta}^* }_{nn^{\prime}}({\bf k})
j^{x(y)}_{nn^{\prime}}({\bf k})}{E_{n^\prime {\bf k}} 
- E_{n {\bf k}}}\theta (-E_{n^\prime {\bf k}})\theta (E_{n {\bf k}}). \nonumber\\
\end{eqnarray}

In the following, Hund's coupling $J$ is set to be
$U$/4 unless stated otherwise~\cite{schafgans},
whereas intraorbital Coulomb interaction is chosen so
that the total magnetic moment per site is nearly unity
for the undoped case.  

\section{Results}

\subsection{Three band model}

Fig.~\ref{daghofer1}(a) and (b) show the mean-field magnetic 
order parameter and the charge density of each of the orbitals in the 
($\pi$, 0)-SDW state for $U = 1.22$eV. Total magnetic moment 
increases continuously on moving from the electron- to hole-overdoped 
region. The difference in magnetization $m_{xz}-m_{yz}$ of two 
orbitals $d_{xz}$ and $d_{yz}$ reverses it's sign slightly 
below band-filling $n = 4.0$ which
corresponds to the undoped case. A similar but opposite trend is 
observed for the ferro-orbital order $n_{xz}-n_{yz}$. 
As can also be seen from Fig.~\ref{daghofer1}(d), sign reversal 
of the orbital order is not related to the reversal of anisotropy
$R_{an} = D_x$/$D_y -1$, where $D_x$ and $D_y$ are Drude weights along
$x$- and $y$-directions, respectively. Latter occurs near hole
doping $x_h \approx 0.22$ in agreement with the experiments. Note that SDW state
for $x_h \approx 0.22$ may be stabilized only at higher temperature \cite{avci}.
$R_{an}$ exhibits a maximum near zero doping and it's 
behavior in the electron doped region where it decreases is in
contrast with the experiments. 

Note that $d_{xz}$ DOS remains dominant at
the Fermi level for the entire hole-doping regime as shown
in Fig.~\ref{daghofer1}(c). Thus, suppression of $d_{yz}$ DOS
at Fermi level alone is not sufficient to explain the sign
reversal. Apart from $d_{xz}$, $d_{xy}$ orbital also plays a
very important role in the anisotropy. As seen from Table I,
interorbital $D^{13}$ and intraorbital $D^{33}$ contributions
to the Drude weight are significant since 
the intraorbital hopping parameters $t_{13}$ and $t_{33}$ are
considerably larger than $t_{11}$ or $t_{22}$, where superscript
1, 2, and 3 have been used for the orbitals $d_{xz}$, $d_{yz}$,
and $d_{xy}$, respectively. $D^{13}_{x}$ remains larger than
$D^{13}_{y}$ despite a reduction in it's value for the entire
hole doping regime considered. At the same time, $D^{33}_{y}$
becomes larger than $D^{33}_{x}$ and contributes, apart from 
small contributions from $D^{11}_{y}$ and $D^{22}_{y}$, 
substantially to the sign-reversal.

\begin{table}[t]
\caption{Components of Drude weight in the three orbital 
model for $U$ = 1.22eV. 
$\alpha$ and $\beta$ are orbital 
indices having values 1, 2, and 3 corresponding to the 
$d_{xz}$, $d_{yz}$, and $d_{xy}$ orbitals, respectively.
Subscripts 1, 2 and 3 of the components correspond 
to  $x_h$ = 0.0, $x_h = 0.24$ and $x_e = 0.1$
respectively.} 
\centering 
\setlength{\tabcolsep}{3pt} 
    \renewcommand{\arraystretch}{1} 
\begin{tabular}{c|c c c c c c c c c c c c c c c c} 
\hline\hline 
$\alpha \beta$ & $11$ & $12$ & $13$ & $22$ & $23$ & $33$  \\ [0.5ex] 
\hline
$D^{\alpha \beta}_{x;1}$ &$0.010$ &  $0.000$ &  $0.098$ & $-0.001$ &  $0.006$ &  $0.067$ \\
\hline
$D^{\alpha \beta}_{y;1}$ &$0.016$ &  $0.000$ & $-0.010$ &  $0.001$ &  $0.007$ &  $0.070$ \\
\hline
$D^{\alpha \beta}_{x;2}$ &$0.002$ &  $0.000$ &  $0.045$ &  $0.000$ &  $0.017$ &  $0.017$ \\
\hline
$D^{\alpha \beta}_{y;2}$ &$0.011$ &  $0.000$ &  $0.019$ &  $0.012$ &  $0.012$ &  $0.041$ \\
\hline
$D^{\alpha \beta}_{x;3}$ &$0.002$  &$ 0.001$ &$ 0.117$ &$ -0.001$ &$ 0.006$ &$ 0.074$ \\
\hline
$D^{\alpha \beta}_{y;3}$ &$0.009$ &$ 0.000$&$ -0.014$ &$ 0.002$ &$ 0.012$ &$ 0.104$ \\
[1ex] 
\hline \hline 
\end{tabular}
\end{table} 

\begin{figure}[b]
\begin{center}
\psfig{figure=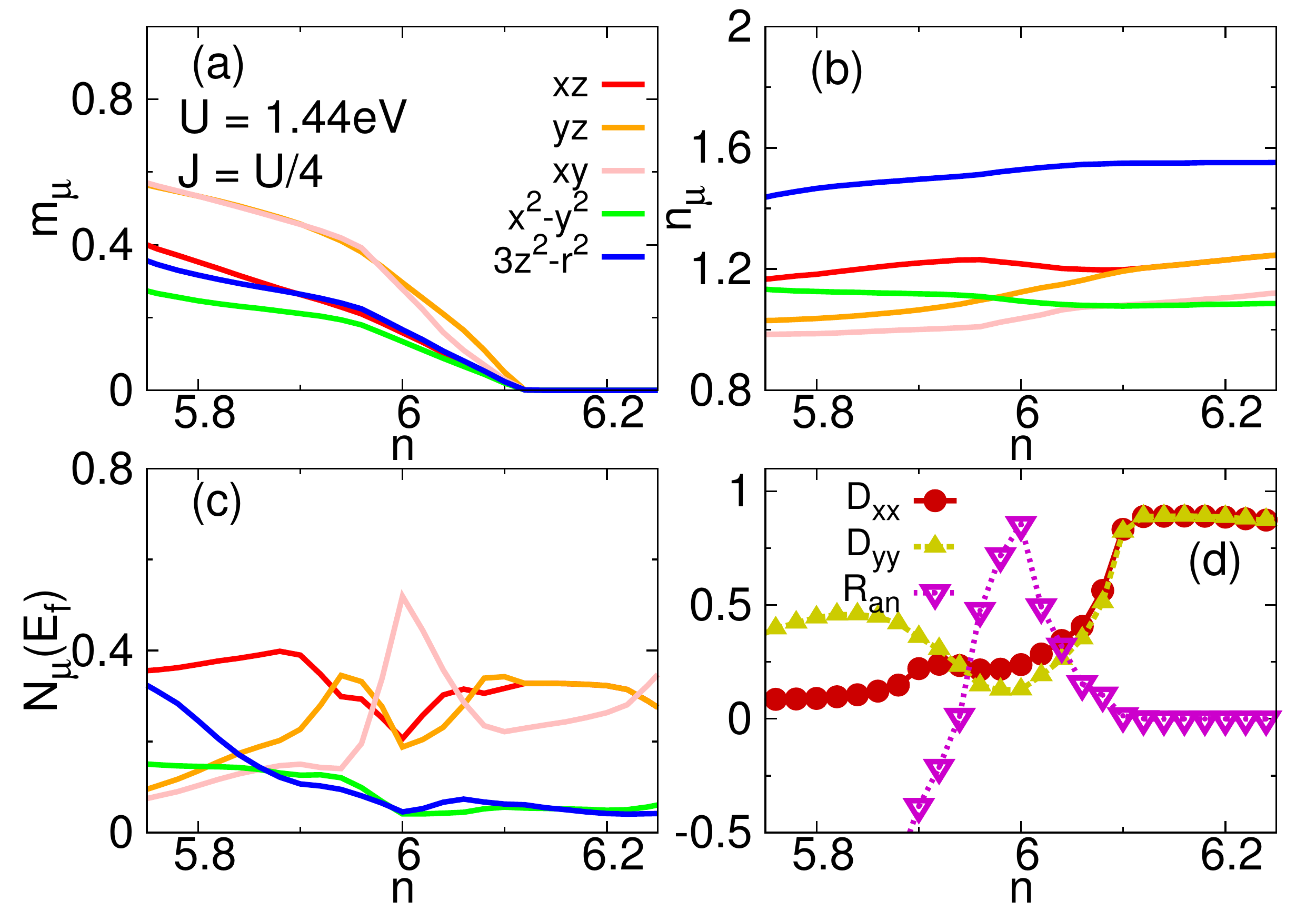,width=6.2cm,height=8cm,angle=-90}
\end{center}
\vspace{-4mm}
\caption{(a) Orbital magnetizations (b) orbital densities
(c) orbital resolved DOS at the Fermi level as a function of doping
in the five-orbital model of Graser \textit{et. al.}. (d)
Drude weight in the $x$-direction, $y$-direction, and the anisotropy. }
\label{graser1}
\end{figure}
\begin{figure}[t]
\begin{center}
\psfig{figure=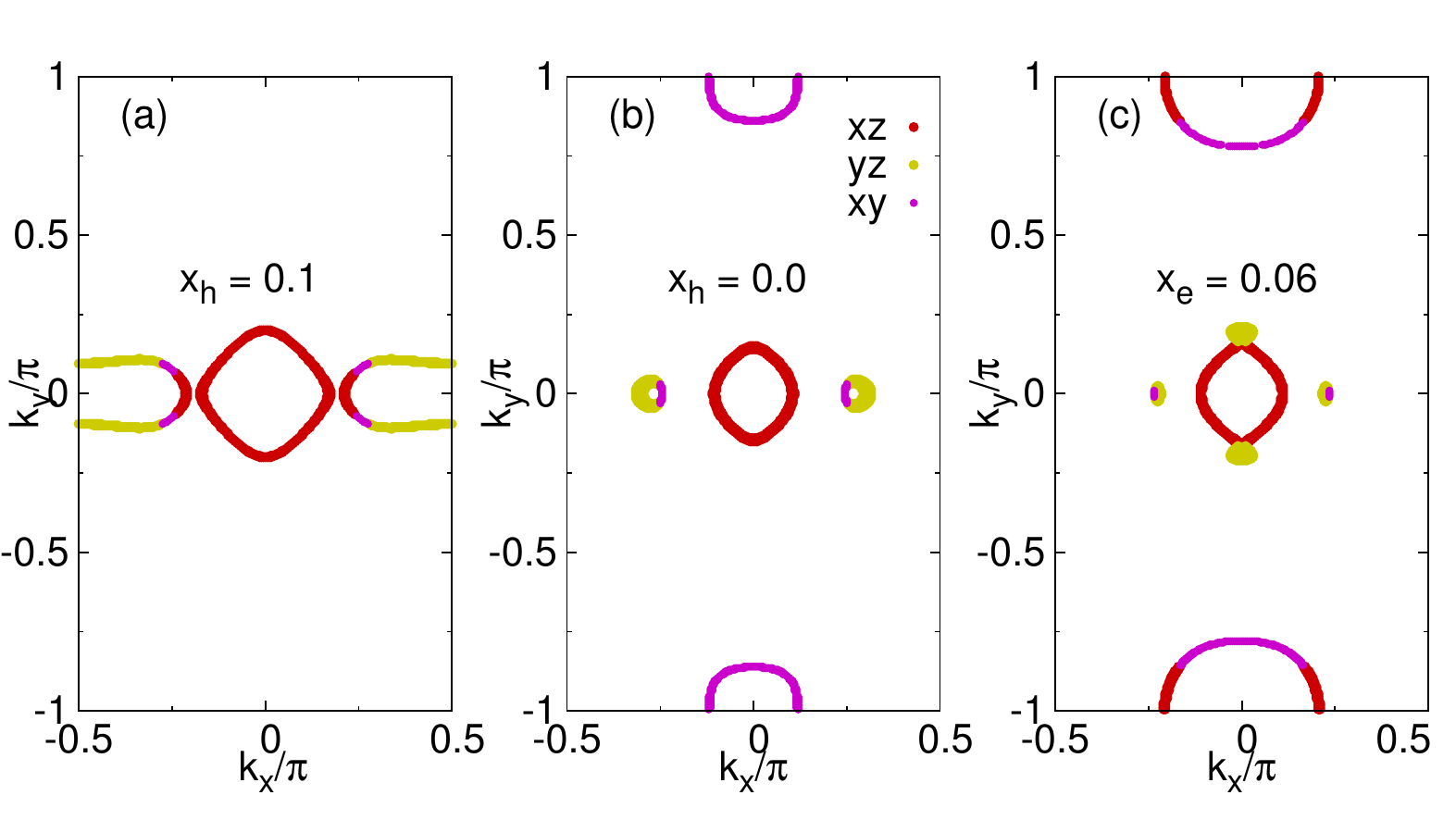,width=80mm,height=5.6cm,angle=0}
\end{center}
\vspace{-4mm}
\caption{Fermi surfaces for hole dopings (a) $x_h = 0.1$, (b)  $x_h = 0.0$, 
and electron doping (c)  $x_e = 0.06$  with dominant orbital character in 
the model of Graser \textit{et. al}., where $U = 1.44$eV. }
\label{graser2}
\end{figure}

To understand the doping dependence
it is essential to take into account the orbital-resolved DOS 
across the Fermi surface as well as the geometrical 
structure of the Fermi surfaces. Reconstruction of the hole pocket~\cite{daghofer,yi1,lu,liu} 
around $\Gamma$ is not seen (Fig.~\ref{daghofer2}(b)) which 
is in contrast with the experiments. This happens
primarily because of larger 
interaction parameters chosen to yield magnetic 
moment $m_{tot} \approx 1$. However, 
reconstruction of the hole pockets for
same magnetization can be clearly seen
in the five-orbital model as presented below, 
which can disappear again on increasing the interaction~\cite{kovacic}.

The contribution to 
$D^{13}_{x(y)}$ comes from those regions of the Fermi surfaces,
which have significant proportions of both the orbitals 
$d_{xz}$ and $d_{yz}$. In the undoped case, such regions are near 
($\pm 0.24\pi,\pm \pi$) of the electron pockets at
($0, \pm \pi$) and also near the boundary 
separating $d_{yz}$ and $d_{xy}$ dominated regions of the hole pockets 
at ($0, \pm0.5\pi$) (Fig.\ref{daghofer2}(b)). Moreover,
components of the Fermi velocity along those regions follow the 
relation $v^{f}_x$ $>$ $v^{f}_y$. For this reason, 
$D^{13}_{x}$ is significantly larger than $D^{13}_{y}$, and thus 
primarily responsible for the Drude weight 
anisotropy. Other important contributor $D^{33}_{x(y)}$
do not differ substantially in the $x$- and $y$-directions for zero doping. 
This is because nearly equal regions of the electron 
pocket at ($0, \pm \pi$) and the hole pockets 
at ($\pm 0.5\pi, 0$) dominated by the $d_{xy}$ 
orbital have Fermi velocity components mainly along 
$y$- and $x$-direction, respectively. The
hole pocket around ($0, 0$) is nearly circular and 
predominantly of $d_{xz}$ character, thus does
not contribute to the anisotropy. 

On electron doping, $d_{xy}$ dominated sections of
the electron pockets and 
hole pockets get shrunk and enlarged, respectively,
 (Fig.~\ref{daghofer2}(c)). This leads to $D^{33}_{y}
> D^{33}_{x}$ causing a drop in $R_{an}$. Whereas 
contribution due to $D^{13}_{(x)y}$ becomes smaller
(Fig.~\ref{daghofer2}(a)) on hole doping. Finally, the regions
of square-like Fermi surface with dominant $d_{xz}$ and $d_{xy}$ orbitals 
have $v^f_y > v^f_x$ in large region leading to $D^{33}_{y} > D^{33}_{x}$, 
which results in the reversal of Drude-weight anisotropy.

\subsection{Five band model of Graser et al.}

In the five-orbital model of Graser \textit{et. al.},
orbital magnetization shows a quick rise on hole doping
so that $\sum_{\mu} m_{\mu} \ge 1.6$ as soon as doping 
reaches 0.1, while it decreases rapidly on electron
doping and vanishes near $n \approx 6.12$ 
(Fig.~\ref{graser1}(a)). $n \approx 6.0$
corresponds to the undoped case $x = 0$.
There is a significant difference in the
magnetizations as well as in the charge densities of 
$d_{xz}$ and $d_{yz}$ orbitals. The ferro-orbital
order $n_{xz} - n_{yz}$ has a sign opposite to that of 
$m_{xz} - m_{yz}$ in the entire region of 
electron density considered here (Fig.~\ref{graser1}(b)).
Unlike the three-orbital model, there is no sign reversal of these
parameters.

Before discussing the Drude weight anisotropy
in this model, it is interesting to look first 
at the orbitally-resolved relative
DOS at the Fermi level. As seen from 
Fig.~\ref{graser1}(c), the DOS of $d_{xy}$ orbital is
the largest for the undoped case in contrast 
with the three-orbital model, which decreases upon doping either 
holes or electrons. On the other hand, DOSs 
of $d_{xz}$ and $d_{yz}$ orbitals, though small and nearly
equal for zero doping, increase on doping either electrons or holes.   

\begin{table}[t]
\caption{Elements of Drude weight in the 
model of Graser \textit{et. al.} for $U = 1.44$eV. Here,
orbital indices having values 1, 2, 3, 4,
and 5, which correspond to the  $d_{xz}$, $d_{yz}$, 
$d_{x^2-y^2}$, $d_{xy}$, and $d_{3z^2-r^2}$ orbitals, respectively. 
 Subscripts 1, 2 and 3
 correspond to $x_h$ = 0.0, $x_h = 0.1$ and $x_e = 0.06$, respectively} 
\centering 
\setlength{\tabcolsep}{2.5pt} 
    \renewcommand{\arraystretch}{1} 
\begin{tabular}{c|c c c c c c c c c c c c c c c c} 
\hline\hline 
$\alpha \beta$ & $11$ & $14$ & $23$ &  $24$ & $35$ & $44$ & $45$ \\ [0.5ex] 
\hline
$D^{\alpha \beta}_{x;1}$ & $0.017$ &  $0.037$  & $0.002$  &  $0.000$ &  $-0.001$ &  $0.019$ &  $0.000$ \\
\hline
$D^{\alpha \beta}_{y;1}$ & $0.010$ &  $0.000$  & $0.001$  &  $0.004$  &  $0.000$ &  $0.017$ &  $0.000$ \\
\hline
$D^{\alpha \beta}_{x;2}$ & $0.016$ &  $0.035$  & $0.000$  &  $0.000$ &  $0.000$ &  $0.024$ &  $0.000$   \\
\hline
$D^{\alpha \beta}_{y;2}$ & $0.010$ &  $-0.001$ & $0.021$  &  $0.043$  &   $0.008$ &  $0.003$ &  $0.019$  \\
\hline
$D^{\alpha \beta}_{x;3}$ & $ 0.030$   & $ 0.085$  & $   -0.002$ & $ 0.000$ & $ 0.000$ & $ 0.010$ & $ 0.000$\\
\hline
$D^{\alpha \beta}_{y;3}$ & $0.007$   & $  0.003$ & $  0.000$ & $ 0.021$ & $ 0.000$ & $ 0.069$ & $ 0.000$\\    
[1ex] 
\hline \hline
\end{tabular}
\end{table} 

\begin{figure}[b]
\begin{center}
\psfig{figure=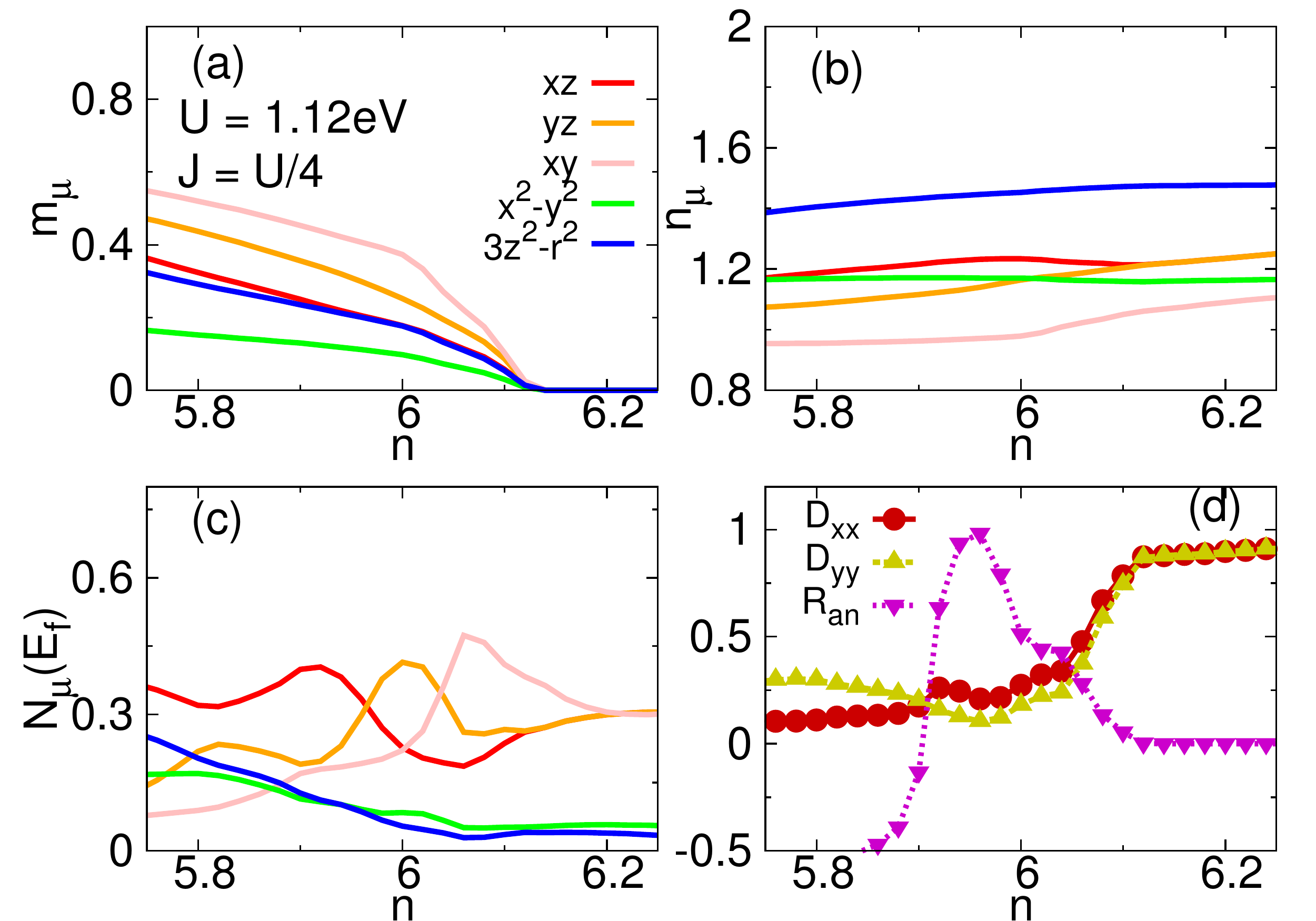,width=6.2cm,height=8cm,angle=-90}
\end{center}
\vspace{-4mm}
\caption{(a) Magnetization, (b) charge density (c)
DOS at the Fermi level for each of the orbitals.
(d) Drude weights in $x$- and $y$-directions, and
their ratios in the model of Ikeda \textit{et. al}..}
\label{ikeda1}
\end{figure}
\begin{figure}[t]
\begin{center}
\psfig{figure=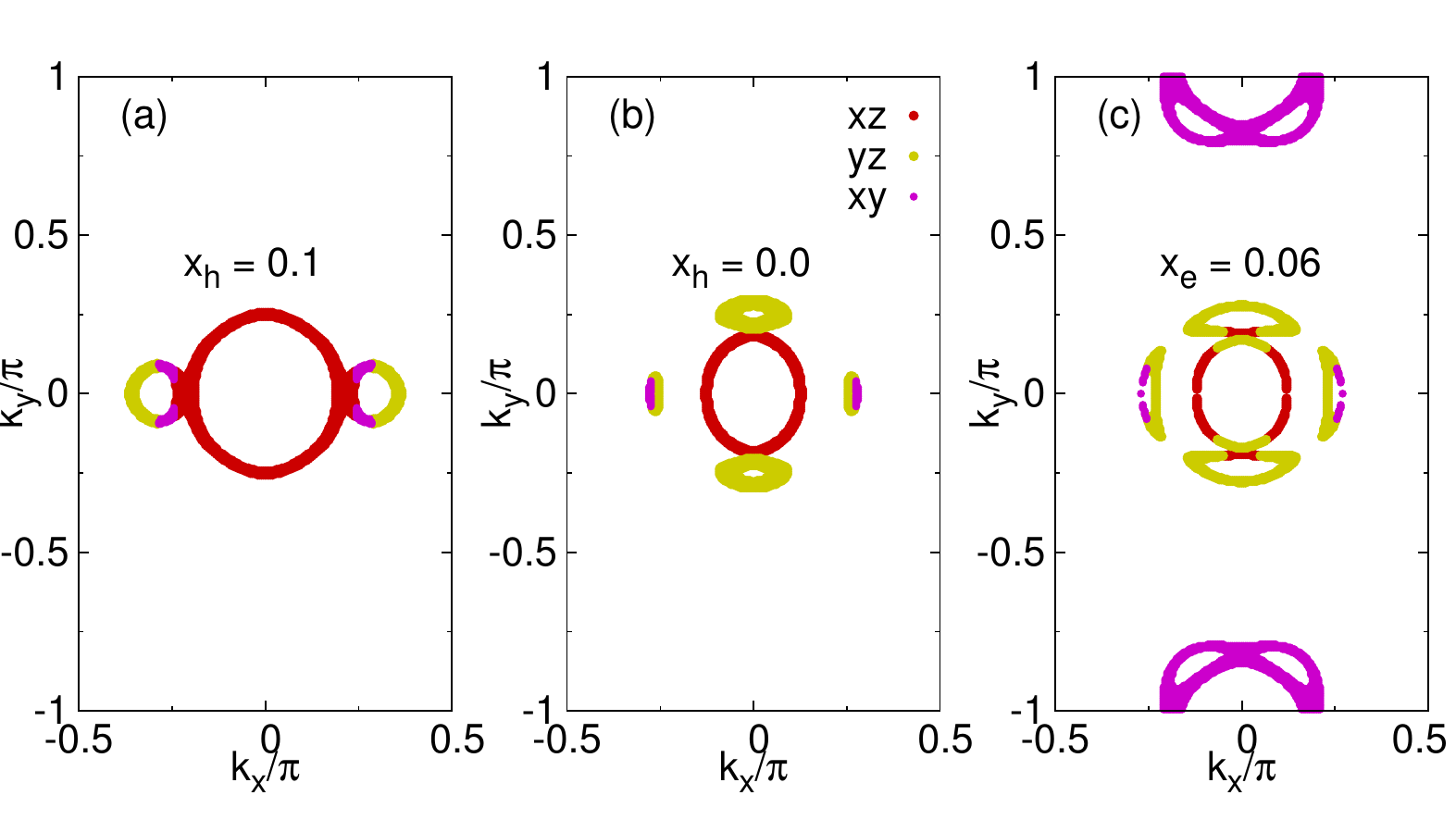,width=80mm,height=5.6cm,angle=0}
\end{center}
\vspace{-4mm}
\caption{Fermi surfaces for hole dopings (a)
$x_h = 0.1$, (b) $x_h = 0.0$, and 
(c) $x_e = 0.06$ with color showing leading
orbital character in the five-orbital model of
Ikeda \textit{et. al.}, where $U = 1.12$eV. }
\label{ikeda2}
\end{figure}

$R_{an}$ is maximum near $x_h \approx 0$, drops rapidly,
and changes sign near $x_h \approx 0.07$ (Fig.~\ref{graser1}(d)). 
This drop is significantly faster in comparison 
with the three-orbital model. Another important difference
from the previous model is that the Drude
weight is larger in the $x$-direction 
despite DOSs of $d_{xz}$ and $d_{yz}$ orbitals being 
nearly equal at the Fermi level in the undoped case. 
Although the same component $D^{14}$ is responsible
for this anisotropy (superscript 4 here refers to
$d_{xy}$ orbital, Table II), which implies again 
a crucial role of the orbitally-resolved DOS distribution along 
the Fermi surfaces. The sections of the Fermi 
surfaces having significant proportions of both 
$d_{xz}$ and $d_{xy}$ orbitals are near ($\pm 0.12\pi, \pm \pi$)
of the electron pockets at ($0, \pm \pi$) and are also near 
the $d_{xy}$ dominated regions of the small hole pockets 
at ($\pm 0.27 \pi, 0$) (Fig.~\ref{graser2}(b)).
Moreover, $v^f_x > v^f_y$ in these regions, 
which is mainly responsible for the anisotropy. 
Like the three-orbital model, the hole pocket at (0, 0) is predominantly 
of $d_{xz}$ orbital character ($\approx$ 88\%),
thus does not play an important role in the anisotropy.

\begin{figure*}
\begin{center}
\psfig{figure=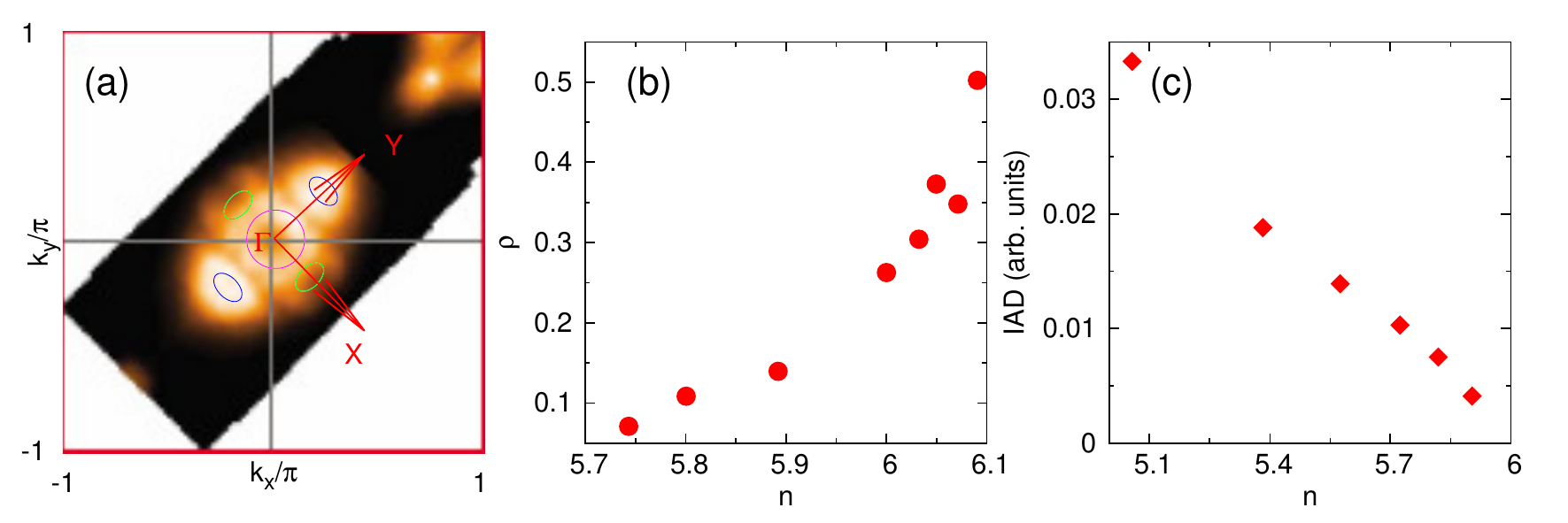,width=160mm,height=4.8cm,angle=0}
\end{center}
\vspace{-4mm}
\caption{(a) ARPES Fermi surfaces obtained for
3.5$\%$ electron-doped CaFe$_2$As$_2$ in 
the orthorhombic state with collinear
magnetic order. $k_x$ and $
k_y$ are the reciprocal lattice vectors defined
corresponding to the unit cell consisting of two Fe atoms (Ref.
[34]). High
symmetry directions in the SDW state are
rotated by 45$^{\circ}$. (b) Resistivity anisotropy
$\rho = \rho_{b}/\rho_{a} - 1$ in BaFe$_2$As$_2$
at 0.95$T_{\rm N}$ reproduced from Ref. [7], where $b$ and $a$ are the
ferromagnetic and the antiferromagnetic axes,
respectively. (c) Integrated absolute difference (IAD) from X-ray emission
spectroscopy as a function of hole doping
for BaFe$_2$As$_2$, which provides a qualitative estimate of the behavior of
local moments as a function of hole doping (Ref. [41]).}
\label{exp}
\end{figure*}
The electron pocket grows with a significant modification
in the orbitally-resolved DOS distribution on electron doping 
in such a way that $D^{14}_x$ increases further while $D^{14}_y$
remains vanishingly small (Fig.~\ref{graser2}(c)). This is 
because the regions of the pocket with nearly equal
proportions of both $d_{xz}$ and $d_{xy}$ orbitals 
have $v^f_x > v^f_y$ . However, $D^{44}_y$ 
becomes significantly larger than $D^{44}_x$
(table II) as $d_{xy}$ orbital dominates now only
those sections which have $v^f_x < v^f_y$. 
For this reason, there is a net decline in $R_{an}$.

On the other hand, the electron pocket becomes
smaller and disappear on hole doping $x_h \approx 0.1$. Meanwhile, 
the hole pockets at ($\pm 0.27 \pi, 0$) grows 
larger in size until they get connected to the 
hole pocket at (0, 0) and also extended in the 
opposite direction (Fig.~\ref{graser2}(a)). 
As seen from table II, anisotropy in $D^{14}$ remains largely 
unaffected because of the regions of hole pockets
having $v^f_x > v^f_y$ significant proportions of 
$d_{xz}$ and $d_{xy}$ orbitals. Whereas the flat
regions of the hole pocket with $v^f_y > v^f_x$ 
(vanishingly small $v^f_x$) have significant 
proportions of $d_{yz}$ as well as of either $d_{xy}$ or 
$d_{x^2-y^2}$ orbitals, which leads to 
$D^{23}_y > D^{23}_x$ and $D^{24}_y > D^{24}_x$
(Table II), hence leading to the 
sign reversal of $R_{an}$.

\subsection{Five band model of Ikeda et al.}

\begin{table}[b]
\caption{Elements of Drude weight in the model
of Ikeda \textit{et. al.} for $U = 1.12$ eV. Here,
orbital indices having values 1, 2, 3, 4, and 5,
which correspond to the  $d_{xz}$, $d_{yz}$, $d_{x^2-y^2}$,
$d_{xy}$, and $d_{3z^2-r^2}$ orbitals, respectively.
 Subscripts 1, 2 and 3
 correspond to $x_h$ = 0.0, $x_h = 0.1$ and $x_e = 0.06$, respectively}
\centering 
\setlength{\tabcolsep}{2.5pt} 
    \renewcommand{\arraystretch}{1} 
\begin{tabular}{c|c c c c c c c c c c c c c c c c} 
\hline\hline 
$\alpha \beta$ &$11$ & $14$  &  $23$  & $24$ & $35$ &$44$ & $45$ \\ [0.5ex] 
\hline
$D^{\alpha \beta}_{x;1}$ & $0.026$&  $0.038$  &   $-0.002$ &  $0.000$ &  $-0.001$ &$0.010$ &  $0.000$ \\
\hline
$D^{\alpha \beta}_{y;1}$ &$0.012$ &  $-0.001$  &  $0.000$  &  $0.014$ &  $0.005$ & $0.000$ & $0.000$ \\
\hline
$D^{\alpha \beta}_{x;2}$ & $0.027$&  $0.021$   &  $0.002$ &  $0.000$ &  $-0.004$ & $0.006$ &  $-0.002$   \\
\hline
$D^{\alpha \beta}_{y;2}$ & $0.013$&  $-0.001$  &  $0.003$  &   $0.024$ &  $0.011$ & $0.001$& $0.006$  \\
\hline
$D^{\alpha \beta}_{x;3}$ & $0.025$& $ 0.057$  & $ 0.003$ & $0.001$ & $ 0.002$ &$0.034$ & $ -0.001$\\
\hline
$D^{\alpha \beta}_{y;3}$ & $0.013$ & $0.001$  & $ 0.000$ & $0.032$ & $ 0.005$ &$0.024$ & $ 0.000$\\
[1ex] 
\hline \hline
\end{tabular}
\end{table}

\begin{table}[b]
\caption{Components of Drude weight in the model
of Daghofer \textit{et. al.} for $\delta = 80$meV. 
$\alpha$ and $\beta$ are orbital indices having
values 1, 2, and 3 corresponding 
to the $d_{xz}$, $d_{yz}$, and $d_{xy}$ orbitals,
respectively.} 
\centering 
\setlength{\tabcolsep}{3pt} 
    \renewcommand{\arraystretch}{1} 
\begin{tabular}{c|c c c c c c c c c c c c c c c c} 
\hline\hline 
$\alpha \beta$ & $11$ & $12$ & $13$ & $22$ & $23$ & $33$  \\ [0.5ex] 
\hline
$D^{\alpha \beta}_{x}$ &$0.008$ &  $0.000$ &  $0.136$ & $0.029$ &  $-0.003$ &  $0.099$ \\
\hline
$D^{\alpha \beta}_{y}$ &$0.013$ &  $-0.001$ & $-0.010$ &  $0.034$ &  $0.135$ &  $0.121$ \\
[1ex] 
\hline\hline 
\end{tabular}
\end{table} 
\begin{table}[t]
\caption{Elements of Drude weight in the model of
Graser \textit{et. al.} for $\delta = 80$meV. Here,
orbital indices having values 1, 2, 3, 4, and 5,
which correspond to the
$d_{xz}$, $d_{yz}$, $d_{x^2-y^2}$, $d_{xy}$, 
and $d_{3z^2-r^2}$ orbitals, respectively.} 
\centering 
\setlength{\tabcolsep}{2.5pt} 
    \renewcommand{\arraystretch}{1} 
\begin{tabular}{c|c c c c c c c c c c c c c c c c} 
\hline\hline 
$\alpha \beta$ & $11$ & $22$ & $33$ &  $44$ & $55$ & $13$ & $14$ \\ [0.5ex] 
\hline
$D^{\alpha \beta}_{x}$ & $0.043$ &  $0.024$  & $0.001$  &  $0.063$ &  $0.000$ &  $0.007$ &  $0.100$ \\
\hline
$D^{\alpha \beta}_{y}$ & $0.023$ &  $0.045$  & $0.002$  &  $0.080$  &  $0.000$ &  $0.026$ &  $0.004$ \\
[1ex] 
\hline\hline 
\end{tabular}
\end{table} 
\begin{table}[t]
\caption{Elements of Drude weight in the model of 
Ikeda \textit{et. al.} with parameters and conventions as 
in Table V.} 
\centering 
\setlength{\tabcolsep}{2.5pt} 
    \renewcommand{\arraystretch}{1} 
\begin{tabular}{c|c c c c c c c c c c c c c c c c} 
\hline\hline 
$\alpha \beta$ & $11$ & $22$ & $33$ &  $44$ & $55$ & $13$ & $14$ \\ [0.5ex] 
\hline
$D^{\alpha \beta}_{x}$ & $0.050$ &  $0.020$  & $0.004$  &  $0.071$ &  $0.000$ &  $0.007$ &  $0.096$ \\
\hline
$D^{\alpha \beta}_{y}$ & $0.020$ &  $0.050$  & $0.005$  &  $0.092$  &  $0.000$ &  $0.033$ &  $0.004$ \\
[1ex] 
\hline\hline 
\end{tabular}
\end{table} 

The doping dependence of magnetic moment and charge 
density of each of the orbitals is similar to that 
of the model of Graser \textit{et. al.} (Fig.~\ref{ikeda1}(a) 
and (b)). However, 
the DOS of $d_{yz}$ orbital is largest at
the Fermi level in the undoped case in 
contrast with the previously discussed two 
models (Fig.~\ref{ikeda1}(c)). 
Behavior of $R_{an}$ as a function of 
hole doping is more or less similar to
the model of Graser \textit{et. al.} except that 
the peak and zero of the anisotropy are shifted 
towards a point near $x_h \approx 0.04$ and $x_h \approx 0.09$,
respectively (Fig.~\ref{ikeda1}(d)). 
Another significant difference from the 
previous model is the absence of the electron pocket 
around ($0, \pm\pi$) for zero doping.

The leading source of anisotropy for 
the undoped case is $D^{14}_x > D^{14}_y$, which originates 
mainly from the small electron pockets near
($\pm 0.25\pi, 0$), which are extended 
along $y$-direction. The reason is similar
to that described in the previous model in terms of the orbitally-resolved 
DOS distributions (Fig.~\ref{ikeda2}(b)). The 
anisotropy vanishes close to $x_h \approx 0.1$
because of the hole pockets being 
nearly circular (Fig.~\ref{ikeda2}(b)). Major components 
contributing to the Drude weight within the range 
$0 \lesssim x_h \lesssim 0.1$ remains similar to that within the model of
Graser {\it et. al.}. Moreover, their behavior as a function of 
doping is also similar. But there are important differences in the electron-doped
region. Here, the suppression of anisotropy does not 
result from a large anisotropy in the $D^{44}$ component ($D_y^{44} >> D_x^{44}$). In fact, 
$D_x^{44}$ remains greater than $D_y^{44}$ largely because of the difference
in orbital-weight distribution
along the electron pocket near ($\pm\pi, 0$). Thus, unlike $D^{44}$,
$D^{24}$ is mainly responsible for the reduction in anisotropy on electron doping.

\subsection{Anisotropy in the nematic phase}

In order to understand the conductivity
anisotropy in the nematic phase, which is found to 
be significant $\sigma_a/\sigma_b \sim 1.2$ 
in the transport measurements, we consider 
an explicit orbital-splitting 
term as observed by the ARPES measurements 
in addition to the kinetic part of the Hamiltonian (Eq. (1)) 
\begin{equation}
\mathcal{H}_{OO} = \frac{\delta}{2}\sum_{i} 
(d^{\dagger}_{{\bf i} yz \sigma} d_{{\bf i} yz \sigma} - 
d^{\dagger}_{{\bf i} xz \sigma} d_{{\bf i} xz \sigma}).
\end{equation}

Consideration of the term has limited motivation
that is to focus on it's impact on the anisotropy instead
of it's origin. Recent works trace the origin to many-body 
effects beyond the random-phase approximation level \cite{onari}. Furthermore, an 
alternative scenario of conductivity anisotropy, where electrons gets scattered by 
the critical-spin fluctuations, has also been proposed~\cite{fernandes}.

Table IV, V and VI show relevant
components of Drude weight obtained for $\delta = 80$meV in
the undoped case. Interestingly, we find anisotropy
in all the three models to be arising due to the
anisotropy in the interorbital component of Drude
weight corresponding to the $d_{xy}$ orbital
indicating that the orbital splitting affects
orbital distribution of $d_{xy}$ the most along
the Fermi surfaces. Anisotropy is
significant $\sim$ 8$\%$ ($D_y > D_x$ ) in each
model but smaller against experimentally
observed $\sim$ 20$\%$. Thus, orbital splitting
induced redistribution of $d_{xy}$ orbital may contribute
significantly to the conductivity anisotropy in the nematic phase.

\section{Discussion}
The ratio of the Drude weights along 
the antiferromagnetic and ferromagnetic directions drops from
$D_x/D_y \sim 2$ to a lower value on moving
away from $x \sim 0$. This compares well with  
with $\sigma_x/\sigma_y \sim 1.5$ for 
$\omega \rightarrow 0$ and $x \sim 0$ in the optical
conductivity measurement where it is found to 
be $\sim 1.5$~\cite{nakajima}. A similar value of 
conductivity ratio is obtained in the  transport
measurements highlighting the important role of 
band structure.

Despite the similarities exhibited by these models with respect to 
doping dependence of local moment and Drude weight
anisotropy, several features of their Fermi surface 
are different. In the models of Daghofer \textit{ et. al.}
and Graser \textit{et. al.}, a very important role in the
anisotropy is played by the electron pockets at ($0, \pm \pi$),
absent in the model of Ikeda \textit{et. al.}. ARPES measurements
do find a weak signature of small electron pockets around X. However, most crucial
features such as the electron pockets placed slightly away from $\Gamma$ 
along $\Gamma$-X and $\Gamma$-Y of the model of Ikeda \textit{et. al.}
are in very good agreement with the experiments except the ellipse-like 
hole pocket around $\Gamma$ which may be either absent or whose 
signature may be very weak to be visible \cite{wang}. In other words,
yellow electron pockets and the electron pocket with two different
regions having different leading orbital character
(Fig.~\ref{ikeda2}(c)) show a good correspondence to the blue and
the green electron pockets of experimental Fermi surfaces
(Fig.~\ref{exp}(a)), respectively.  Overall, a better agreement
with the experiments regarding the Fermi-surface characteristics
is exhibited within this model. It is also important to note
that the electron pockets are mainly responsible for the
anisotropy even in this model. Moreover, similar electron
pockets occur also in the model of Graser \textit{et. al.} though
their role is not as significant as in the model of Ikeda 
\textit{et. al.} because of their smaller sizes.

Drude weight anisotropy in all the three model explored here 
exhibit a common trend. Continuous decline and 
sign reversal on hole doping compares very well the experiments
while the decline on electron doping is in contrast with the 
measurements~\cite{blomberg}. The decrease of anisotropy on hole
doping is not so fast in the three-orbital model
owing largely to a slow modification in the electronic structure,
whereas a sharp decline is exhibited in the five-orbital models.

Magnetic moment grows on moving from the electron-doped towards
the hole-doped region irrespective of the models. This is
not surprising because in each of the models, one 
approaches half filling on doping holes, which
suppresses electron movement in the lattice leading to the
departure from metallicity, and finally to the Mottness as 
also pointed out by a recent density functional theory (DFT)
combined with slave-spin mean-field approach~\cite{medici}. Evidence in 
support of growing magnetic moment on hole doping has been 
substantiated recently by the X-ray emission spectroscopy using
integrated absolute difference (IAD)~\cite{lafuerza}.

\begin{figure}[t]
\vspace{-0mm}
\begin{center}
\psfig{figure=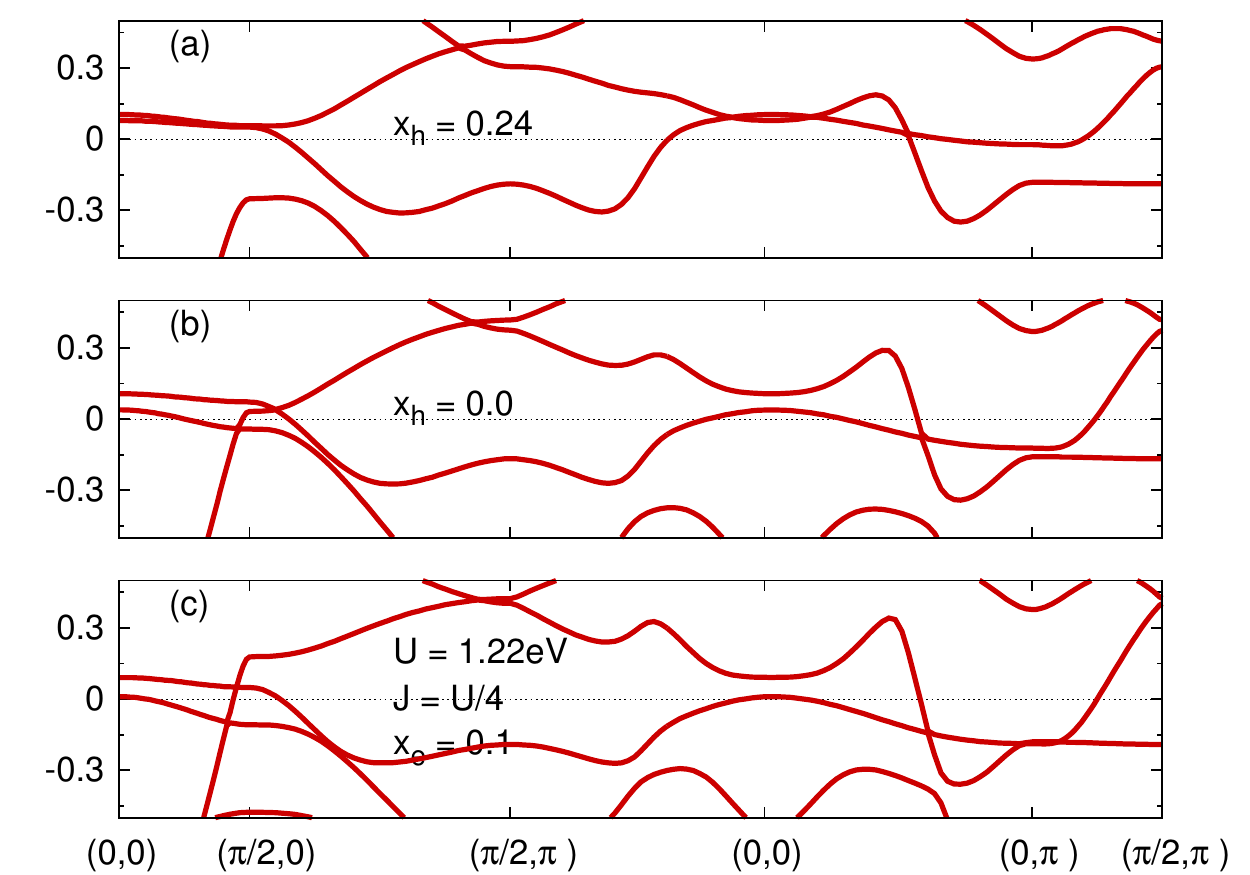,width=85mm,angle=0}
\end{center}
\vspace{-5mm}
\caption{Electronic dispersion in the ($\pi, 0$)
-SDW state along high-symmetry directions
within the three-orbital model of Daghofer \textit{et. al.}. }
\label{disp1}
\end{figure}
\begin{figure}[]
\vspace{-0mm}
\begin{center}
\psfig{figure=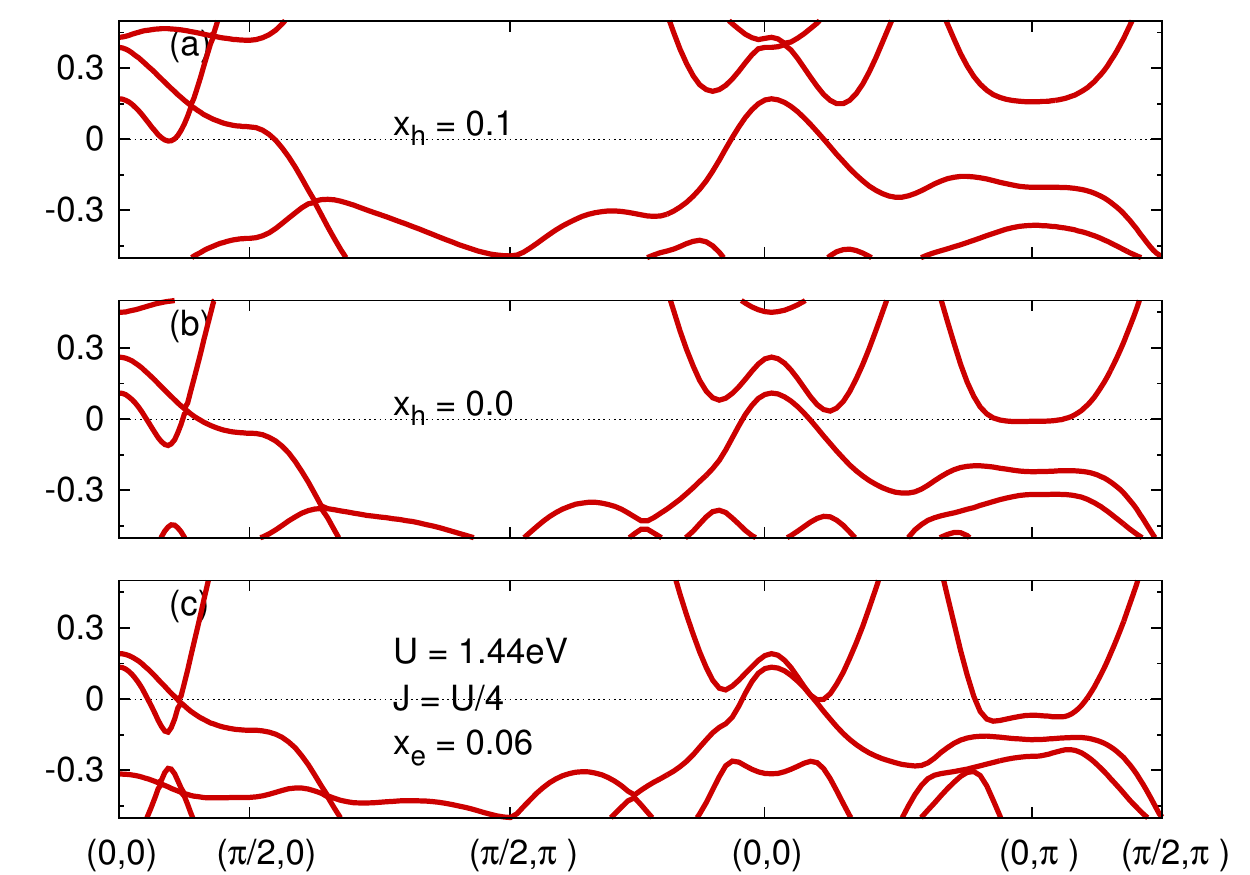,width=85mm,angle=0}
\end{center}
\vspace{-5mm}
\caption{Same as in Fig.\ref{disp1} but for the five-orbital model
of Graser \textit{et. al.}.}
\label{disp2}
\end{figure}
\begin{figure}[b]
\vspace{-0mm}
\begin{center}
\psfig{figure=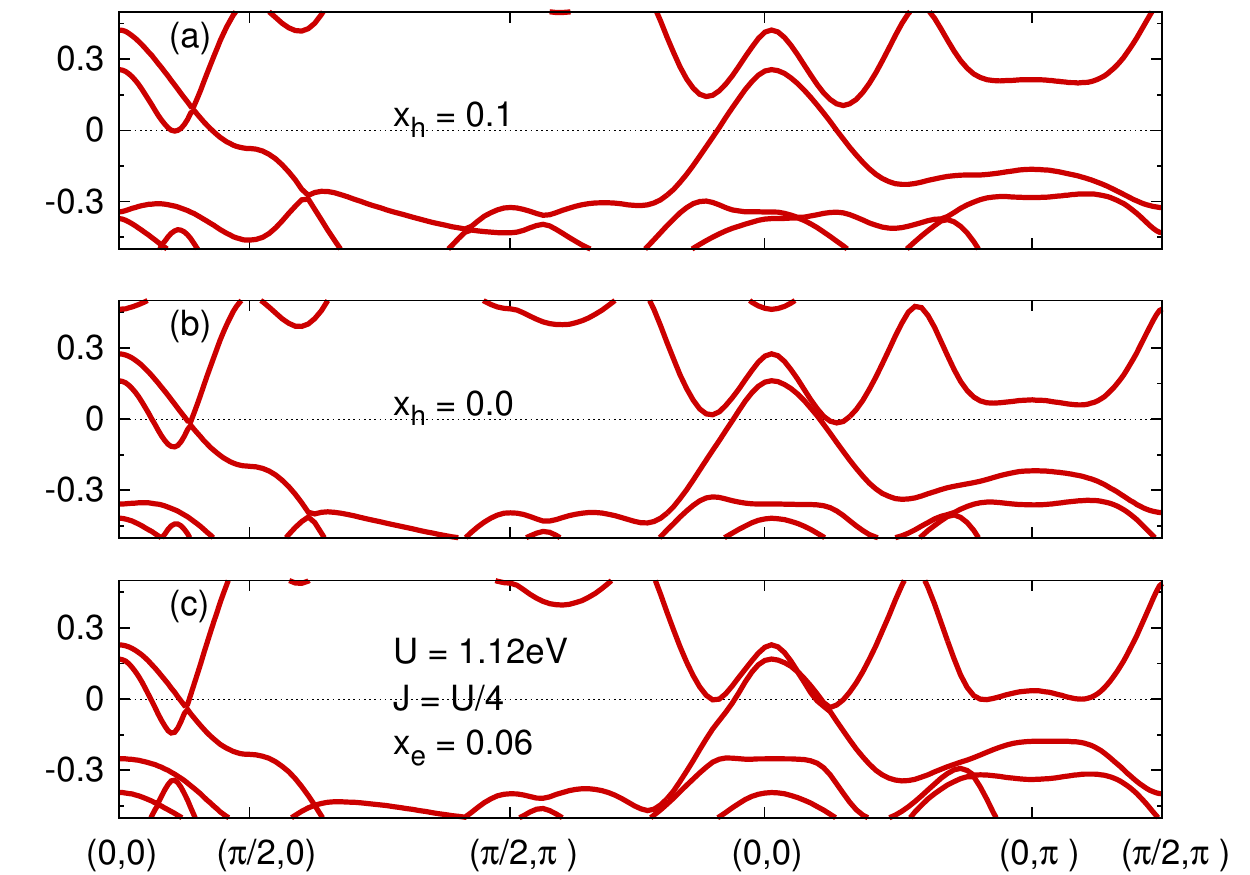,width=85mm,angle=0}
\end{center}
\vspace{-5mm}
\caption{Same as in
Fig.\ref{disp2} but for the model of Ikeda \textit{et. al.}.}
\label{disp3}
\end{figure}
\section{conclusions} 
We have established the role of bandstructure 
on the doping dependence of conductivity anisotropy 
by examining the Drude weights within the widely used models of pnictides. 
In all of them, the anisotropy exhibits a 
maximum near zero doping, decreases on doping holes, and then
reverses sign at a certain hole doping, all in qualitative 
agreement with the experiments. However, we also find that 
the anisotropy decreases on electron doping and vanishes, a feature in 
contrast with the experiments, clearly implying 
the role of other factors such as 
doping induced disorder.

We explain the origin of anisotropy and it's 
doping dependence 
in terms of the topology and shape of the 
Fermi surfaces as well as the distribution of
orbital-resolved DOS along them - not just the dominance of 
some particular orbital at the Fermi level. This is 
reflected in the similar anisotropy behavior of different models 
despite the dominance of different low energy orbitals 
in them.  The leading contributor 
to the anisotropy common for all the models is 
interorbital hopping processes involving 
$d_{xz}$ and $d_{xy}$ orbitals. 
In the nematic phase, a 
significant anisotropy is obtained due to the redistribution of 
orbital weight which affects $d_{xy}$ orbital the most. Overall, the 
$d_{xy}$ orbital plays the most 
crucial role in the conductivity anisotropy in the iron pnictides.

\section*{Acknowledgment}

We acknowledge use of the HPC Clusters at HRI.

\section*{Appendix: Electronic dispersion in the ($\pi, 0$)-SDW state}

Electronic dispersions for the different models are presented for the parameters
discussed above. Fig.~\ref{disp1} shows dispersion along high-symmetry 
directions for the dopings (a) $x_h = 0.24$, (b) $x_h = 0.0$, and (c) $x_e = 0.1$
in the three-orbital model of Daghofer \textit{et. al.} with $U = 1.22$eV. 
Fig.~\ref{disp2} 
shows dispersion for the dopings (a) $x_h = 0.1$, (b)
$x_h = 0.0$, and (c) $x_e = 0.06$ in the five-orbital model of Graser \textit{et. al.} 
with $U = 1.44$eV. 
Fig.~\ref{disp3} shows dispersion for the dopings (a) $x_h = 0.1$, (b) $x_h = 0.0$, 
and (c) $x_e = 0.06$ in the five-orbital model of Ikeda \textit{et. al.} with $U = 1.12$eV.

\end{document}